\documentclass{article}
\usepackage{spconf, amsmath, epsfig, amssymb, booktabs, soul, color, xcolor}

\usepackage{hyperref}
\hypersetup{
    colorlinks=true,
    linkcolor=blue,
    urlcolor=red,
    citecolor=blue,
}

\title{Sparse Focus Network for Multi-Source Remote Sensing Data Classification}

\name{Xuepeng Jin$^{1}$, Junyan Lin$^{1}$, Feng Gao$^{1}$, Lin Qi$^{1}$ Yang Zhou$^{2}$}
\address{$^1$ School of Computer Science and Technology, Ocean University of China, Qingdao 266100, China \\
$^2$ China Electronic Standardization Institute Huadong Branch, Suzhou 215124, China
\thanks{This work was supported in part by the National Key Research and Development Program of China under Grant 2022ZD0117202 and in part by the Natural Science Foundation of Qingdao under Grant 23-2-1-222-ZYYD-JCH. (Corresponding author: Feng Gao, Email: gaofeng@ouc.edu.cn)}}

\begin{document}
\maketitle

\begin{abstract}

Multi-source remote sensing data classification has emerged as a prominent research topic with the advancement of various sensors. Existing multi-source data classification methods are susceptible to irrelevant information interference during multi-source feature extraction and fusion. To solve this issue, we propose a sparse focus network for multi-source data classification. Sparse attention is employed in Transformer block for HSI and SAR/LiDAR feature extraction, thereby the most useful self-attention values are maintained for better feature aggregation. Furthermore, cross-attention is used to enhance multi-source feature interactions, and further improves the efficiency of cross-modal feature fusion. Experimental results on the Berlin and Houston2018 datasets highlight the effectiveness of SF-Net, outperforming existing state-of-the-art methods.

\end{abstract}


\section{Introduction}

With the continuous development of various satellite sensors, multi-source remote sensing image classification has gradually become one of the hot research topics \cite{zhao2022joint}. Multi-source remote sensing data classification refers to the process of labeling pixels using data from different sensors, such as hyperspectral image (HSI), synthetic aperture radar (SAR), and light detection and ranging (LiDAR) \cite{wang2020adaptive}. Multi-source data classification has a wide range of applications in disaster assessment \cite{wang22grsl}, urban planning \cite{cheng21igarss} and vegetation monitoring \cite{xu22igarss}.

In this paper, we mainly focus on HSI-SAR/LiDAR data classification. HSI provides rich spectral information, SAR sensor is sensitive to the shape and structure of ground objects, and LiDAR provides detailed height information of ground objects. Recently, many deep learning-based methods have been proposed for multi-source data classification. By introducing attention mechanisms \cite{wm23tgrs, mohla2020fusatnet}. Wang et al. \cite{wm23tgrs} employed bilinear attention fusion for HSI and LiDAR data joint classification. Mohla et al. \cite{mohla2020fusatnet} used the self-attention and cross-attention for multi-source remote sensing data classification. These methods automatically learn the important information from the input data, thereby improving model performance.  Although attention mechanisms have achieved excellant performance in multi-source data classification, they still have some limitations, such as being susceptible to irrelevant information interference during information interaction. 

To address the issue, this study introduces a sparse attention mechanism to optimize the model's feature interaction process. Sparse attention reduces interference from irrelevant features, retaining only the most relevant self-attention values. To this end, we proposed a \textbf{S}parse \textbf{F}ocus \textbf{Net}work (SF-Net) for HSI-SAR/LiDAR data classification. Specifically, sparse attention is employed in Transformer block for HSI and SAR/LiDAR feature extraction, thereby the most useful self-attention values are maintained for better feature aggregation. In addition, cross-attention is used to enhance multi-source feature interactions, and further improves the efficiency of cross-modal feature fusion. Extensive experimental results on the Berlin and Houston2018 datasets demonstrate the superior classification performance over existing state-of-the-art methods.

\begin{figure*}[h]
\centering
\includegraphics [width=6.5in]{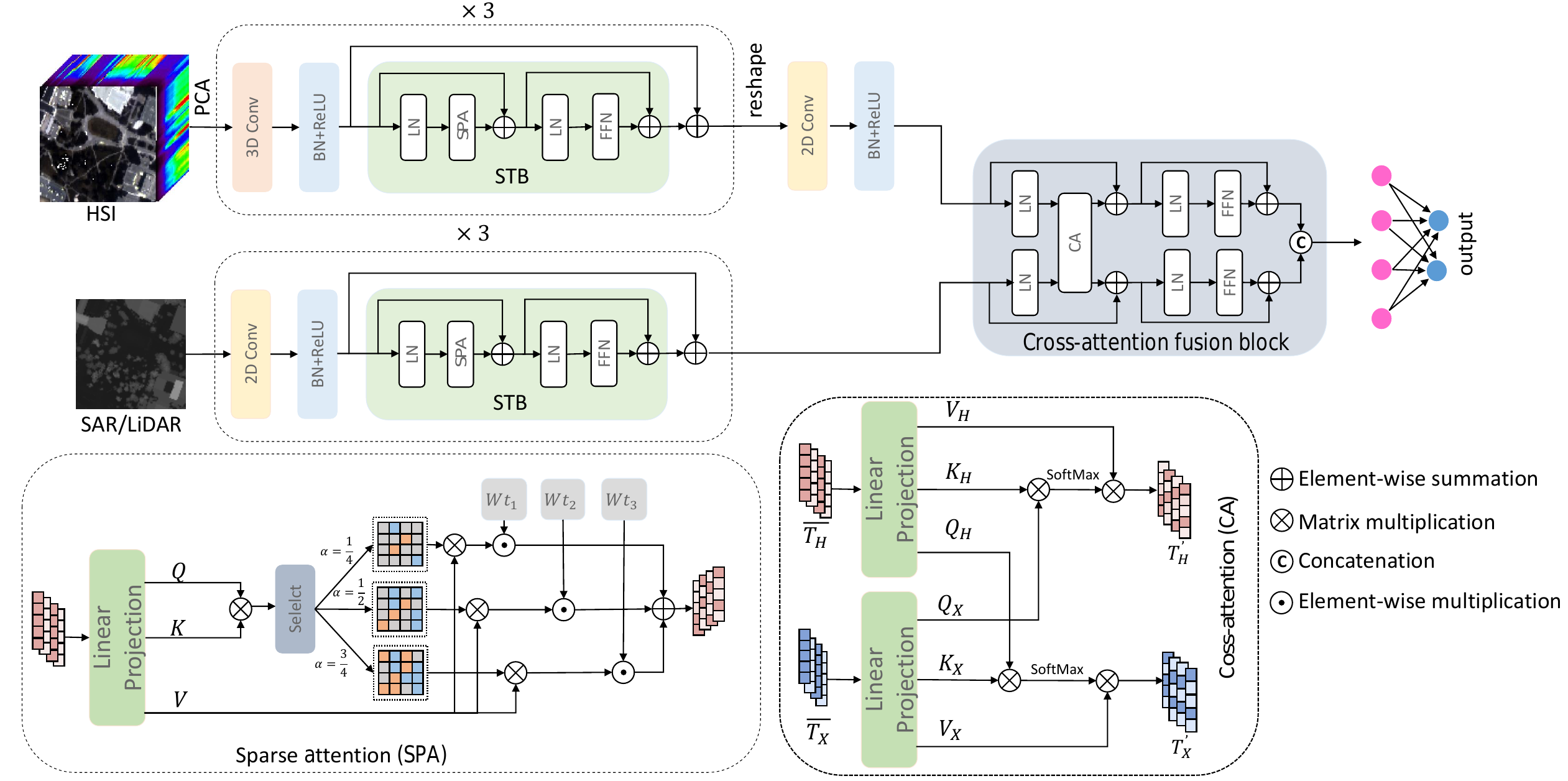}
\caption{Overview of the proposed SF-Net for HSI and LiDAR/SAR classification.}
\label{fig_framework} 
\end{figure*}

\begin{figure*}[h]
\centering
\includegraphics [width=6.5in]{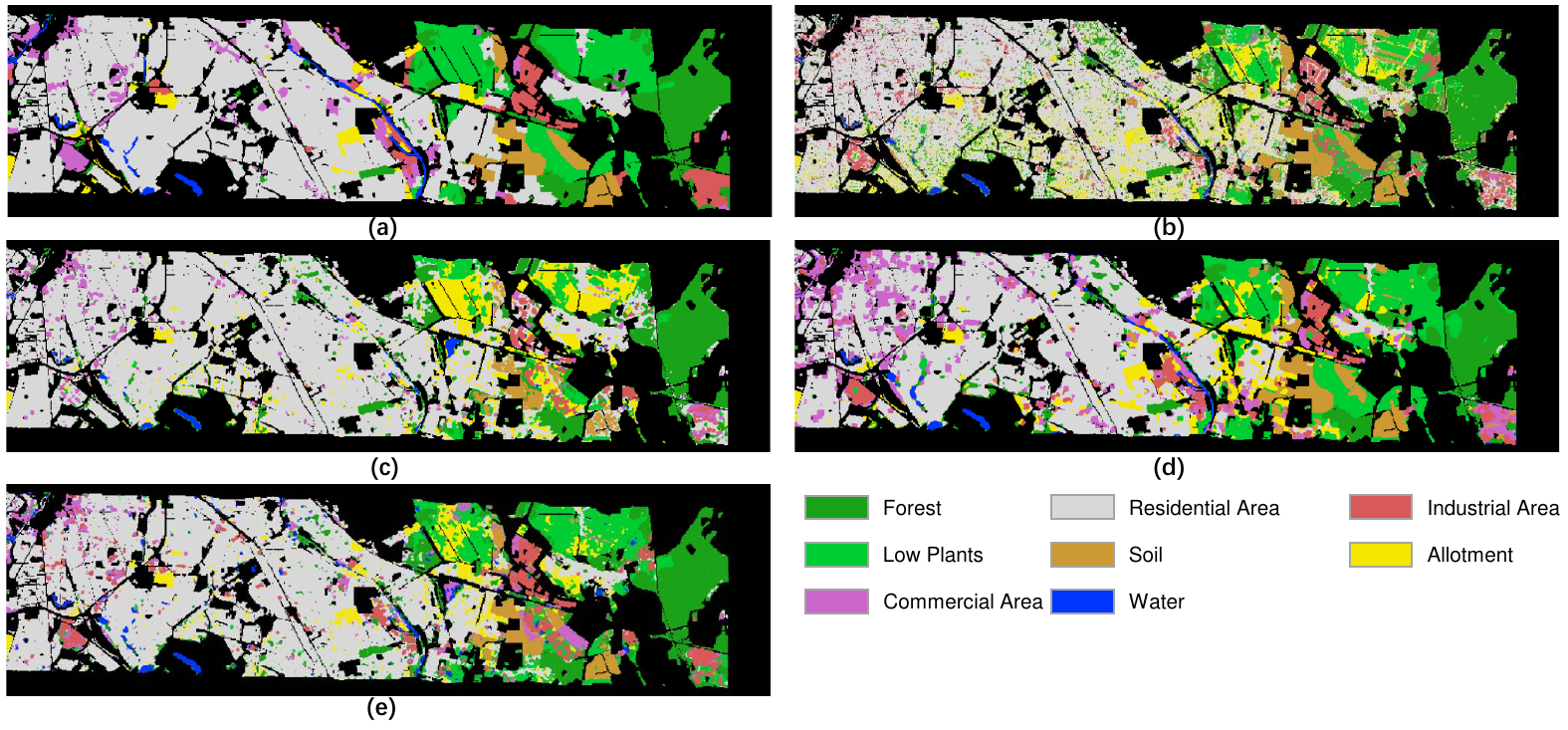}
\caption{Classification results of different methods for the Berlin dataset. (a) The ground truth. (b) TBCNN. (c) FusAtNet. (d) ExViT. (e) The proposed SF-Net.}
\label{result} 
\end{figure*}

\section{Methodology}

Our proposed SF-Net framework for the classification of HSI and SAR/LiDAR is depicted in Figure \ref{fig_framework}. SF-Net consists of the feature extraction module and cross-attention fusion module. For HSI feature extraction, PCA is first employed for spectral-dimension reduction, and 3D convolution is employed. For SAR/LiDAR feature extraction, 2D convolution is employed. The critical part in feature extraction is the Sparse Transformer Block (STB), and the STB is repeated three times. Finally, features from HSI and SAR/LiDAR are fed into the Cross-Attention Fusion Block (CAFB). The outputs of CAFB are transformed into the classification results via fully connected layer.

It should be noted that the structure of STB is similar to the standard Transformer block \cite{dosovitskiy2020image}. The main difference between the STB and standard Transformer block is that STB  utilizes sparse attention.

\subsection{Sparse Attention}

The sparse attention's purpose is to dynamically retain crucial attention scores for each query, boosting feature aggregation efficiency. This mechanism achieves dynamic selection and sparsity in self-attention values, reducing interference from irrelevant information and optimizing the model's performance in multi-source remote sensing image classification.

Fig. \ref{fig_framework} illustrates the details of sparse attention. Given an input matrix $X \in \mathbb{R}^{N \times D}$, where $N$ represents the number of tokens, and $D$ denotes the dimensionality of each token, we utilize three fully connected layers to derive the query $Q \in \mathbb{R}^{N \times D}$, key $K \in \mathbb{R}^{N \times D}$, and value $V \in \mathbb{R}^{N \times D}$. 

Afterward, we perform matrix multiplication on $Q$ and $K$, followed by a selection operation. For each attention matrix $M_\gamma$, the parameter $k_\gamma$ is determined by the following formula:

\begin{equation}
k_\gamma = \lfloor \alpha \cdot N \rfloor,
\end{equation}
where $\alpha$ takes values of 1/2, 2/3, 3/4, 4/5, and $n$ is set to 4. This formula ensures a varied sparsity level for each $M_\gamma$, where $\gamma$ ranges from 1 to $n$. 

The selection operation retains the top $k_\gamma$ largest values while setting the remaining values to 0. Consequently, this process results in the creation of multiple attention matrices with distinct sparsity levels, collectively denoted as $M_\gamma \in \mathbb{R}^{N \times N}$,  Each $M_\gamma$ within the set captures unique attention patterns. To elaborate further, the acquisition of $M_\gamma$ is detailed as follows:

\begin{equation}
\text{Score}=\frac{(QK^{T})}{\sqrt{D}}
\end{equation}
\begin{equation}
M_{\gamma,ij}' = 
\begin{cases}
  \text{Score}_{ij}, & \text{if } j \text{ is in the top }  k_\gamma \text{ values}  \\
  -\infty, & \text{if } \text{otherwise}
\end{cases}
\end{equation}
\begin{equation}
M_{\gamma,i} = \text{softmax}(M_{\gamma,i}')
\end{equation}
where the softmax function transforms the values in each row of $M_{\gamma,i}'$ into a probability distribution. 

For each $M_\gamma$, the subsequent step entails performing matrix multiplication between the acquired $M_\gamma$ and the value matrix $V$. Finally, generate the output result $Z$. The detailed process is outlined as follows:
\begin{equation}
Z=\sum_{\gamma=1}^{n}Wt_\gamma \cdot M_{\gamma}V
\end{equation}
where $Wt_\gamma$ is a learnable weight parameter.

\subsection{Cross-Attention Feature Fusion}

In order to attain more accurate classification results, it is crucial to effectively fuse multi-source features. The thoughtful design of feature interactions allows the multi-source features to complement each other effectively, thereby improving the discriminative capability for ground objects and ultimately achieving superior classification performance. Consequently, we utilize the cross-attention \cite{kim2022cross} mechanism to facilitate the fusion of features between both modalities.

The details of the Cross-Attention Fusion Block (CAFB) is shown in Fig. \ref{fig_framework}. Given the input hsi feature matrix $T_\textrm{H} \in \mathbb{R}^{N \times D}$ and LiDAR/SAR features matrix $T_\textrm{X} \in \mathbb{R}^{N \times D}$, the entire execution process of CB is as follows.
\begin{equation}
T_H',T_X' = \textrm{CA}(\textrm{LN}(T_H),\textrm{LN}(T_X))
\end{equation}
\begin{equation}
T_H''= \textrm{FFN}(\textrm{LN}(T_H+T_H'))+(T_H+T_H')
\end{equation}
\begin{equation}
T_X''= \textrm{FFN}(\textrm{LN}(T_X+T_X'))+(T_H+T_H')
\end{equation}
\begin{equation}
T_\textrm{out} = \textrm{Concat}(T_H'',T_X'')
\end{equation}
where $T_H'$ and $T_X'$ serve as residuals to $T_H$ and $T_X$, respectively, to enhance the model's performance and learning capability. $T_\textrm{out}$ represents the final output of the entire CAFB.

The details of the Cross Attention (CA) are illustrated in Fig. \ref{fig_framework}. Similar to the standard attention computation, the key difference lies in comparing the query from the $\overline{T}_H$ with the key from $\overline{T}_X$ to obtain attention scores. Subsequently, the values $V_X$ derived from $\overline{T}_X$ are used to produce the final result $T_X'$ based on the obtained attention. The process of obtaining $T_H'$ and $T_X'$ is depicted as follows.
\begin{equation}
T_H'=\textrm{softmax}(\frac{Q_{X}K_{H}^T}{\sqrt{D}})V_{H}
\end{equation}
\begin{equation}
T_X'=\textrm{softmax}(\frac{Q_{H}K_{X}^T}{\sqrt{D}})V_{X}
\end{equation}
where $Q_H$, $K_H$, and $V_H$ are obtained from $T_H$ through three fully connected layers, and $Q_X$, $K_X$, and $V_X$ follow a similar process.

\begin{table}[h]
\centering
\caption{Classification Performance (\%) of Different Models on the Berlin Dataset.}
\vspace{0.5em}
\small
\setlength{\tabcolsep}{2pt} 
\begin{tabular}{c|c c c c}
\toprule
Class & TBCNN & Fusatnet & ExViT & SF-Net \\
\midrule
Forest & 81.75 & 86.24 & 78.01 & 74.75 \\ 
Residential area & 76.26 & 91.38 & 74.05 & 91.30 \\ 
Industrial area & 39.67 & 19.76 & 39.48 & 34.36 \\
Low plants & 49.78 & 20.00 & 84.15 &  65.75 \\ 
Soil & 89.42 & 48.72 & 88.03 & 80.13 \\ 
Allotment & 54.36 & 38.89 & 70.00 & 26.05 \\ 
Commercial area & 4.65 & 18.47 & 38.18 & 28.23 \\ 
Water & 41.93 & 29.61 & 56.41 & 38.24 \\ 
\midrule
OA & 67.60 & 70.91 & 72.63 & 74.89 \\ 
\bottomrule
\end{tabular}
\label{table_berlin}
\end{table}

\begin{table}[h]
\centering
\caption{Classification Performance (\%) of Different Models on the Huston2018 Dataset.}
\vspace{0.5em}
\small
\setlength{\tabcolsep}{2pt} 
\begin{tabular}{c|c c c c}
\toprule
Class & TBCNN & Fusatnet & ExViT & SF-Net \\
\midrule
Health grass & 94.84 & 96.28 & 93.65 & 83.07 \\ 
Stressed grass & 92.60 & 93.45 & 95.44 & 96.07 \\ 
Artificial turf & 100.00 & 100.00 & 100.00 & 100.00 \\
Evergreen trees & 98.80 & 98.33 & 98.52 &  92.78 \\ 
Deciduous trees & 97.12 & 99.11 & 99.25 & 79.83 \\ 
Bare earth & 99.61 & 100.00 & 99.92 & 98.54 \\ 
Water & 100.00 &100.00 &100.00 &99.47   \\
Residential buildings &93.21 &97.90 &96.88 &88.66   \\
Non-residential buildings &91.30 &93.41 &94.87 &99.14 \\
Roads &61.06 &74.56 &80.51 &87.98 \\ 
Sidewalks &75.91 &82.94 &80.27 &76.95 \\
Crosswalks &85.31 &84.65 &96.76 &33.02  \\
Major thoroughfares &72.77 &86.90 &82.11 &89.98 \\
Highways &95.86 &97.59 &84.10 &83.24 \\
Railways &99.78 &99.71 &99.81 &97.25 \\
Paved parking lots &90.74 &98.08 &99.08 &92.68 \\
Unpaved parking lots &100.00 &100.00 &100.00 &98.25 \\
Cars &98.47 &97.11 &97.56 &89.93 \\
Trains &99.90 &99.63 &99.90 &96.97\\
Stadium seats &99.92 &99.93 &99.94 &98.13 \\
\midrule
OA & 86.95 & 91.52 & 91.87 & 92.74 \\ 
\bottomrule
\end{tabular}
\label{table_houston}
\end{table}

\section{Experimental Results and Analysis}

To verify the effectiveness of the proposed model, extensive experiments are carried out on the Berlin dataset and Houston 2018 dataset. Berlin dataset is used for HSI and SAR data classification, encompassing both urban and rural areas in Berlin.
The Berlin dataset comprises HSI and SAR images. The HSI image has a resolution of 797$\times$220 pixels, covering a wavelength range from 400 to 2500 nm with 244 spectral bands. In contrast, the SAR image has a resolution of 1723$\times$476 pixels. Both images were captured in urban and rural areas of Berlin, Germany. The Houston dataset includes HSI and LiDAR images. The HSI image consists of 48 spectral bands covering a wavelength range from 380 to 1050 nm. The LiDAR image contains 3 bands. This dataset spans the campus of the University of Houston and its surrounding urban areas in Houston, Texas, USA. It was part of the 2018 IEEE GRSS Data Fusion Competition.

We compare the proposed SF-Net with three state-of-the-art methods, including TBCNN \cite{xu2017multisource} , FusAtNet \cite{mohla2020fusatnet} and ExViT \cite{yao2023extended}. The OA values of all methods are presented in Table \ref{table_berlin} and \ref{table_houston}. As can be observed that the OA value of the proposed SF-Net reaches 74.89\% on the Berlin dataset. On the Houston2018 dataset, the OA value of the proposed SF-Net reaches 92.74\%, significantly outperforming the other methods. Fig. \ref{result} illustrates the classification results for the Berlin dataset. As can be seen that the result of the proposed SF-Net is the closest to the human-labeled ground truth. Both the quantitative and qualitative assessments demonstrate the exceptional performance of the proposed SF-Net in multi-source remote sensing data classification.

\section{Conclusion}

In this paper, we propose SF-Net for multi-source remote sensing data classification, addressing the challenges associated with information overload and irrelevant feature interference. Sparse attention is employed in Transformer block for HSI and SAR/LiDAR feature extraction, thereby the most useful self-attention values are maintained for better feature aggregation. In addition, cross-attention is used to enhance multi-souce feature interactions, and further improves the efficiency of cross-modal feature fusion. Experimental results on the Houston 2018 and Berlin datasets highlight the effectiveness of SF-Net, outperforming existing state-of-the-art methods.


\begin{thebibliography}{99}

\small

\bibitem{zhao2022joint}
G. Zhao, Q. Ye, L. Sun, Z. Wu, C. Pan, and B. Jeon. ``Joint classification of hyperspectral and LiDAR data using a hierarchical CNN and Transformer," \textit{IEEE Transactions on Geoscience and Remote Sensing}, vol. 61, pp. 1--16, 2023.

\bibitem{wang2020adaptive}
J. Wang, F. Gao, J. Dong, and Q. Du, ``Adaptive dropblock-enhanced generative adversarial networks for hyperspectral image classification," \textit{IEEE Transactions on Geoscience and Remote Sensing}, vol. 59, no. 6, pp. 5040--5053, 2021.

\bibitem{wang22grsl}
H. Wang et al., ``Multi-source remote sensing intelligent characterization technique-based disaster regions detection in high-altitude mountain forest areas," \textit{IEEE Geoscience and Remote Sensing Letters}, vol. 19, pp. 1--5, 2022.

\bibitem{cheng21igarss}
Q. Cheng, Q. Chen, Y. Li and B. Cao, ``Analysis of the influence of sky view factor on urban surface temperature based on multi-source data," In \textit{IGARSS} 2021, pp. 6952--6955.

\bibitem{xu22igarss}
Y. Xu, M. Zhang, E. Yu, Y. Hou, C. Yang and S. Deng, ``Assessing temporal and spatial variations of vegetation degradation in southwest China based on multi-source remote sensing data," In \textit{IGARSS} 2022, pp. 6005--6008.


\bibitem{wm23tgrs}
M. Wang, F. Gao, J. Dong, H. Li, and Q. Du, ``Nearest neighbor-based contrastive learning for hyperspectral and LiDAR data classification," \textit{IEEE Transactions on Geoscience and Remote Sensing}, vol. 61, pp. 1--16, 2023.

\bibitem{mohla2020fusatnet}
S. Mohla, S. Pande, B. Banerjee, and S. Chaudhuri, ``Fusatnet: Dual attention based spectrospatial multi-modal fusion network for hyperspectral and LiDAR classification,"  \textit{CVPR Workshop} 2020, pp. 416-425.

\bibitem{dosovitskiy2020image}
A. Dosovitskiy, L. Beyer, A. Kolesnikov, et al. ``An image is worth 16x16 words: Transformers for image recognition at scale," \textit{ICLR} 2021, pp. 1--12.

\bibitem{kim2022cross}
H. Kim, S. Yu, S. Yuan, and C. Tomasi, ``Cross-attention Transformer for video interpolation," \textit{ACCV} 2022, pp. 320–-337.

\bibitem{xu2017multisource}
X. Xu, W. Li, Q. Ran, Q. Du, L. Gao, and
B. Zhang, ``Multisource remote sensing data classification based on convolutional neural network," \textit{IEEE Transactions on Geoscience and Remote Sensing}, vol. 56, no. 2, pp. 937-–949, 2017.



\bibitem{yao2023extended}
J. Yao, B. Zhang, C. Li, D. Hong, and J.
Chanussot, ``Extended vision Transformer (ExViT) for land use and land cover classification: A multimodal deep learning framework," \textit{IEEE Transactions on Geoscience and Remote Sensing}, vol. 61, pp. 1--15, 2023.

\end{thebibliography}
\end{document}